\documentclass[modern]{aastex63}
\usepackage[margin=1.0in, top=1in, bottom=1in]{geometry}
\usepackage[utf8]{inputenc}
\usepackage{graphicx, subfigure}
\usepackage{color}
\usepackage{amsmath}  
\usepackage{verbatim}     
\usepackage{enumerate}
\usepackage{hyperref}
\usepackage[noabbrev,capitalise]{cleveref}  
\usepackage{natbib}
\usepackage[mathscr]{euscript} 
\usepackage{xspace}
\usepackage[disable]{todonotes}

\usepackage{soul}

\makeatletter
\usepackage{etoolbox}
\patchcmd\H@refstepcounter{\protected@edef}{\protected@xdef}{}{}
\makeatother

\newcommand{\sn}{SN~Ia\xspace}
\newcommand{\sne}{SNe~Ia\xspace}
\newcommand{\romanFull}{\textit{Nancy Grace Roman Space Telescope}\xspace}
\newcommand{\romanST}{\textit{Roman Space Telescope}\xspace}
\newcommand{\rst}{\textit{Roman}\xspace}

\newcommand{\duke}{\affiliation{Department of Physics, Duke University, 120 Science Drive, Durham, NC, 27708, USA}}
\newcommand{\gsfc}{\affiliation{NASA Goddard Space Flight Center, Greenbelt, MD 20771, USA}}
\newcommand{\harvard}{\affiliation{Department of of Astronomy, Harvard University, Cambridge MA 02138}}
\newcommand{\kavlicambridge}{\affiliation{Institute of Astronomy and Kavli Institute for Cosmology, Madingley Road, Cambridge, CB3 0HA, UK}}
\newcommand{\kavlichicago}{\affiliation{Kavli Institute for Cosmological Physics, University of Chicago, Chicago, IL 60637, USA}}
\newcommand{\jhu}{\affiliation{Department of Physics and Astronomy, Johns Hopkins University, Baltimore, MD 21218, USA
}}
\newcommand{\lbnl}{\affiliation{E.O. Lawrence Berkeley National Laboratory, 1 Cyclotron Rd., Berkeley, CA, 94720, USA}}
\newcommand{\moore}{\affiliation{Gordon and Betty Moore Foundation, 1661 Page Mill Road, Palo Alto, CA 94304, USA}}
\newcommand{\rutgers}{\affiliation{Department of Physics and Astronomy, Rutgers the State University of New Jersey, 136 Frelinghuysen Road, Piscataway, NJ 08854, USA}}
\newcommand{\stsci}{\affiliation{Space Telescope Science Institute, 3700 San Martin Drive Baltimore, MD 21218, USA}}
\newcommand{\yale}{\affiliation{Department of Physics, Yale University, New Haven, CT, 06250-8121, USA}}
\newcommand{\ucb}{\affiliation{Department of Physics, University of California Berkeley, 366 LeConte Hall MC 7300, Berkeley, CA, 94720, USA}}

\newcommand{\ucsc}{\affiliation{Department of Astronomy and Astrophysics, University of California, Santa Cruz, CA 95064, USA}}
\newcommand{\uhawaii}{\affiliation{Department of Physics and Astronomy, University of Hawai`i at M{\=a}noa, Honolulu, Hawai`i 96822, USA}}
\newcommand{\umbc}{\affiliation{University of Maryland, Baltimore County, Baltimore, MD 21250, USA}}
\newcommand{\uminnesota}{\affiliation{University of Minnesota, Minnesota Institute for Astrophysics, 116 Church St. SE, Minneapolis, MN 55455, USA}}
\newcommand{\upenn}{\affiliation{Department of Physics and Astronomy, University of Pennsylvania, 209 South 33rd Street, Philadelphia, PA 19104, USA}}
\newcommand{\upitt}{\affiliation{Pittsburgh Particle Physics, Astrophysics, and Cosmology Center (PITT PACC). Physics and Astronomy Department, University of Pittsburgh, Pittsburgh, PA 15260, USA}}
\newcommand{\usouthcarolina}{\affiliation{Department of Physics and Astronomy, University of South Carolina, 712 Main St., Columbia, SC 29208, USA}}
\newcommand{\ipac}{\affiliation{Caltech/IPAC, Mailcode 100-22, Pasadena, CA 91125, USA}}

\begin{document}

\title{A Reference Survey for Supernova Cosmology with the Nancy Grace Roman Space Telescope}
\shorttitle{A Reference Survey for Supernova Cosmology}

\author[0000-0002-1873-8973]{B.~M.~Rose}
\duke
\author{C. Baltay}
\yale
\author[0000-0002-0476-4206]{R.~Hounsell}
\umbc
\gsfc
\author[0000-0002-9946-4635]{P.~Macias}
\ucsc
\author[0000-0001-5402-4647]{David Rubin}
\uhawaii
\stsci
\lbnl
\author[0000-0002-4934-5849]{D. Scolnic}
\duke
\author{G. Aldering}
\lbnl
\ucb
\author[0000-0001-9806-0551]{Ralph~Bohlin}
\stsci
\author[0000-0002-5995-9692]{Mi~Dai}
\jhu
\author[0000-0003-2823-360X]{S. E. Deustua}
\affiliation{Ars Metrologia, Lutherville, MD 21093}
 \author[0000-0002-2445-5275]{R.~J.~Foley}
 \ucsc
\author[0000-0002-6652-9279]{A. Fruchter}
\stsci
\author[0000-0002-1296-6887]{L. Galbany}
\affiliation{Institute of Space Sciences (ICE, CSIC), Campus UAB, Carrer de Can Magrans, s/n, E-08193 Barcelona, Spain.}
\affiliation{Institut d'Estudis Espacials de Catalunya (IEEC), E-08034 Barcelona, Spain.}
\author[0000-0001-8738-6011]{S.~W.~Jha}
\rutgers
\author[0000-0002-6230-0151]{D.~O.~Jones}
\altaffiliation{NASA Einstein Fellow}
\ucsc
\author[0000-0002-7593-8584]{B. A. Joshi}
 \jhu
\author[0000-0003-3142-997X]{Patrick L. Kelly}
\uminnesota
\author{R. Kessler}
\kavlichicago
\author[0000-0002-1966-3942]{Robert P. Kirshner}
\harvard
\moore
\author[0000-0001-9846-4417]{Kaisey~S.~Mandel}
 \kavlicambridge
 \author{S. Perlmutter}
 \lbnl
 \ucb
\author[0000-0002-2361-7201]{J. Pierel}
\stsci
\author{H. Qu}
\upenn
\author{D. Rabinowitz}
\yale
\author{A. Rest}
\stsci
\author[0000-0002-6124-1196]{Adam G. Riess}
\stsci
\jhu
\author[0000-0003-1947-687X]{S. Rodney}
\usouthcarolina                  
\author[0000-0003-2764-7093]{M. Sako}
\upenn
\author[0000-0003-2445-3891]{Matthew R. Siebert}
\ucsc
 \author[0000-0002-7756-4440]{L. Strolger}
 \stsci
\author[0000-0001-7266-930X]{N. Suzuki}
\affiliation{Kavli Institute for the Physics and Mathematics of the Universe (Kavli IPMU, WPI), The University of Tokyo, 5-1-5 Kashiwanoha, Kashiwa, Chiba 277-8583, Japan}
\author{S.~Thorp}
\kavlicambridge
\author[0000-0001-9038-9950]{S.~D.~Van Dyk}
\ipac
\author{K.~Wang}
\duke
\author{S. M. Ward}
\kavlicambridge
\author[0000-0001-7113-1233]{W.~M.~Wood-Vasey}
\upitt
\collaboration{0}{the \rst Supernova Science Investigation Teams\vspace{-1em}}
\correspondingauthor{B. M. Rose}
\email{benjamin.rose@duke.edu}
\shortauthors{the \rst Supernova SITs}

\date{\today}

\section{Introduction}

The \romanFull is the top space-based priority from the 2010 Astronomy Decadal Survey and is scheduled to be launched in the mid 2020s. One of the major goals of the mission is to make a generation-defining measurement of dark energy properties. To do so, the \romanST will utilize its Wide Field Instrument (0.28 square-degree field of view) to conduct a High-latitude Wide Area Survey for weak lensing and large-scale structure studies and a High-latitude Time Domain Survey in order to measure the expansion history of the universe with Type Ia Supernovae (\sne).  

This note presents an initial survey design for the High-latitude Time Domain Survey.
This is not meant to be a final or exhaustive list of all the survey strategy choices, but instead presents one viable path towards achieving the desired precision and accuracy of dark energy measurements using \sne. Furthermore, we note that many of the assumptions in this document represent our current best knowledge and may still change based on further study.

\section{Reference Survey}

\sn cosmology with the \romanFull will use data from the High-latitude Time Domain core community survey. Though this survey's definition is forthcoming, an updated reference survey is needed now for planning, simulations and as a baseline for future optimization.

The two Supernova Science Investigation Teams (SITs) have worked together over the past year to update the current reference survey. Specifically the survey will make use of the Wide Field Instrument (WFI) containing eight broadband photometric filters of which we intend to use six: F062, F087, F106, F129, F158, and F184 (R, Z, Y, J, H, and F respectively). These filters cover the wavelength range $0.62\text{--}2.00~\text{microns}$. The WFI also contains a low-resolution prism (0.75--1.80~microns). We prefer the prism over the higher-resolution grism since the prism has a $\sim$2~mag deeper sensitivity for a one-hour exposure.\footnote{Further details on the WFI hardware can be found at \url{https://roman.gsfc.nasa.gov/science/WFI_technical.html} and \url{https://www.stsci.edu/roman/observatory}.} 

The reference survey presented here preserves some features of the previous strategies \citep{Spergel2015,Hounsell2018}: a total of 6 months observing time spread over 2 years, with 30-hour visits every 5 days. We also assume the availability of 800 externally observed low-redshift ($z<0.1$) \sne. These low-redshift \sne act as an anchor for our cosmological analysis.

The top-level cosmology-requirements for the Supernovae Cosmology project (\rst science requirement SN~2.0.1) will require a $\sim$3~mmag sensitivity level in each of roughly 4 bins over a redshift range of $0.5 < z < 2.0$ (adapted from requirement SN~2.0.3). To do so, we must observe enough \sn with a high enough signal-to-noise ratio (S/N) per observation so that we are dominated by the intrinsic scatter floor (roughly $0.1$~mag and $0.06$~mag for imaging and spectroscopy, respectively).  Furthermore, we exclusively rely on rest-frame optical data to reach this precision, which is observer-frame near infrared (NIR) for the survey's redshift range of interest. Rest-frame NIR observations are a promising complementary route for precision \sn distances, and research is on going.  Below, we detail our strategies to reach our required level of precision, and then discuss whether our strategies satisfy cosmological requirements.

\subsection{Filters \& Tiers}

This reference survey employs a two-tier strategy: wide and deep. Four WFI filters per tier are used to cover the rest-frame optical; for the wide tier these are RZYJ, and for the deep tier they are YJHF. 
Prism spectroscopic data will be collected for a yet-to-be-determined fraction of the survey area. 
Therefore, some fraction of the \sne in the final cosmology analysis will not have spectral data and will require an alternative approach to obtaining the redshift and SN identification.
In this note, we begin with the discussion of the 25\% (by time) prism spectroscopy case in describing the reference survey---we consider this 
to be used for initial future simulations. There are a number of drivers that will influence the final fraction of \sne with prism spectra, however, and alternative strategies are discussed in \cref{sec:alt}.

\subsection{Field Choices}

Repeated observation of only a few square degrees of the sky requires careful consideration for the exact choice of field(s) to observe.  There are practical limitations, scientific choices, and legacy value to consider.
Considerations for a choice of \romanST deep fields are laid out in \citet{Foley2018} and \cite{Koekemoer2019}. 
The main drivers are: 
\begin{enumerate}
\item High ecliptic latitude to minimize zodiacal light, and to reach the \romanST continuous-viewing zone ($\gtrsim \pm 54^{\circ}~\text{off the ecliptic}$) to avoid observational seasons and meet \rst science requirement SN~2.3.4
\item High Galactic latitude to minimize dust extinction
\item Overlap with past, current, and planned wide-area surveys (e.g., Spitzer, the Dark Energy Survey, the Vera C.\ Rubin Observatory)
\item Avoid bright stars
\end{enumerate}
These four constraints lead to choices of deep fields at very high/low declination that are currently inaccessible for telescopes in the opposite hemisphere. We note that if the \rst field of regard is improved---even a few degrees---some particularly attractive fields such as the Chandra Deep Field-South would become accessible. The extinction constraint also limits the possible range of right ascension to only a few hours for each hemisphere.

\begin{figure}
    \centering
    \includegraphics[width=0.85\textwidth]{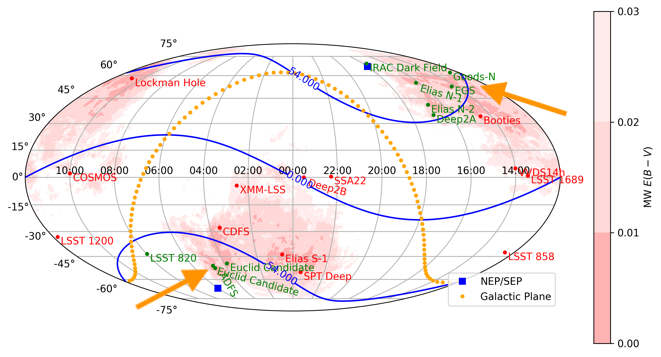}
    \caption{We present possible continuous viewing zone fields (green, $\gtrsim \pm 54^{\circ}~\text{off the ecliptic}$) along with other deep fields (red). Low Milky Way dust extinction is shown as red shading. Overall, the top field choices include GOODS-N (or EGS) and \textit{Euclid} South Deep (indicated with arrows).
    We note that if the field of regard is improved---even a few degrees---some particularly attractive fields such as the Chandra Deep Field-South would become accessible.
    Figure from \citet{Foley2018b}.
    \label{fig:field}
    }
\end{figure}

When choosing the number of fields, a smaller number will improve observing efficiency by minimizing edge losses and slew times as well as having a more internally-consistent calibration. A larger number of fields however will provide more diversity, which can improve follow-up observations, reduce issues related to cosmic variance (and specifically correlation of the \sne gravitational lensing signal), provide tests of isotropy, and potentially aid tests of systematic uncertainties. There is a large gain in having at least two fields, since that allows observations from ground-based telescopes in both hemispheres and will provide a simple ``jackknife'' test.  
We believe two fields---one in the North, and one in the South---is a reasonable choice for a nominal observing strategy. Specific opportunities for fields are presented in  \cref{fig:field}, with top field choices including GOODS-N and \textit{Euclid} South Deep (the AKARI Deep Field South).

\subsection{Slewing Strategy, Roll Angles \& Area}
\label{sec:slew}

\begin{figure}
    \centering
    \includegraphics[width=0.52\textwidth]{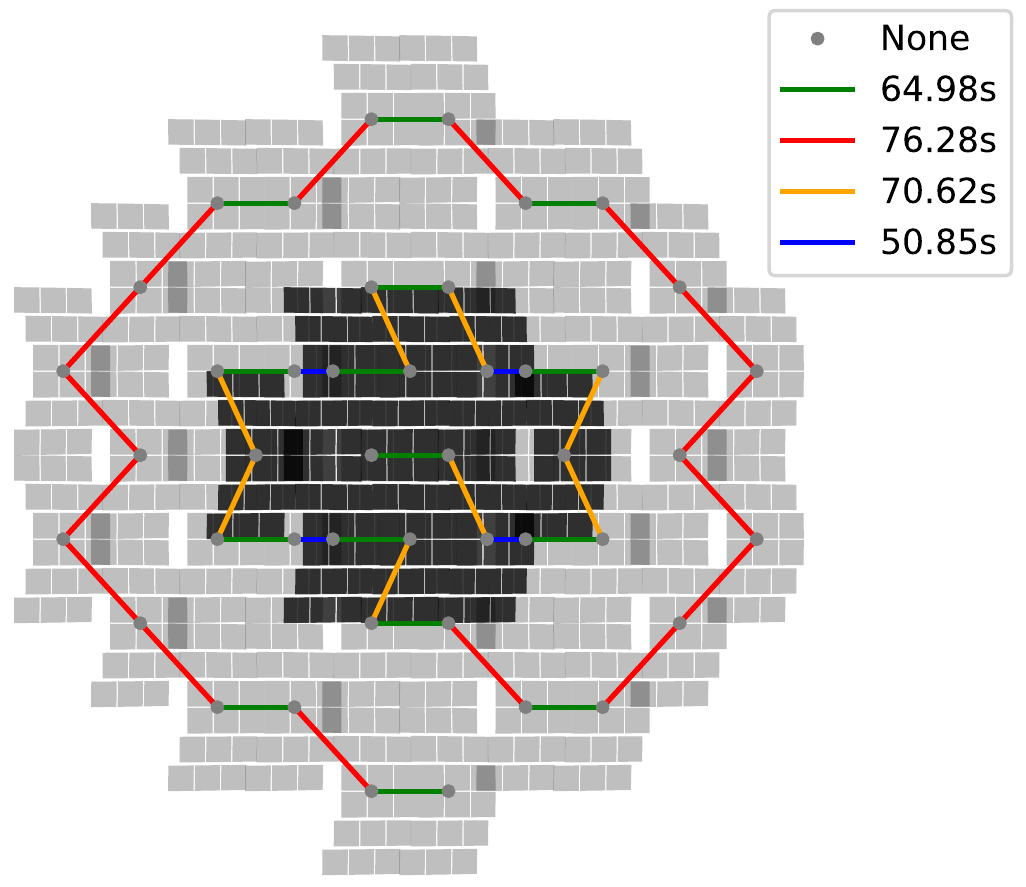}
    \includegraphics[width=0.47\textwidth]{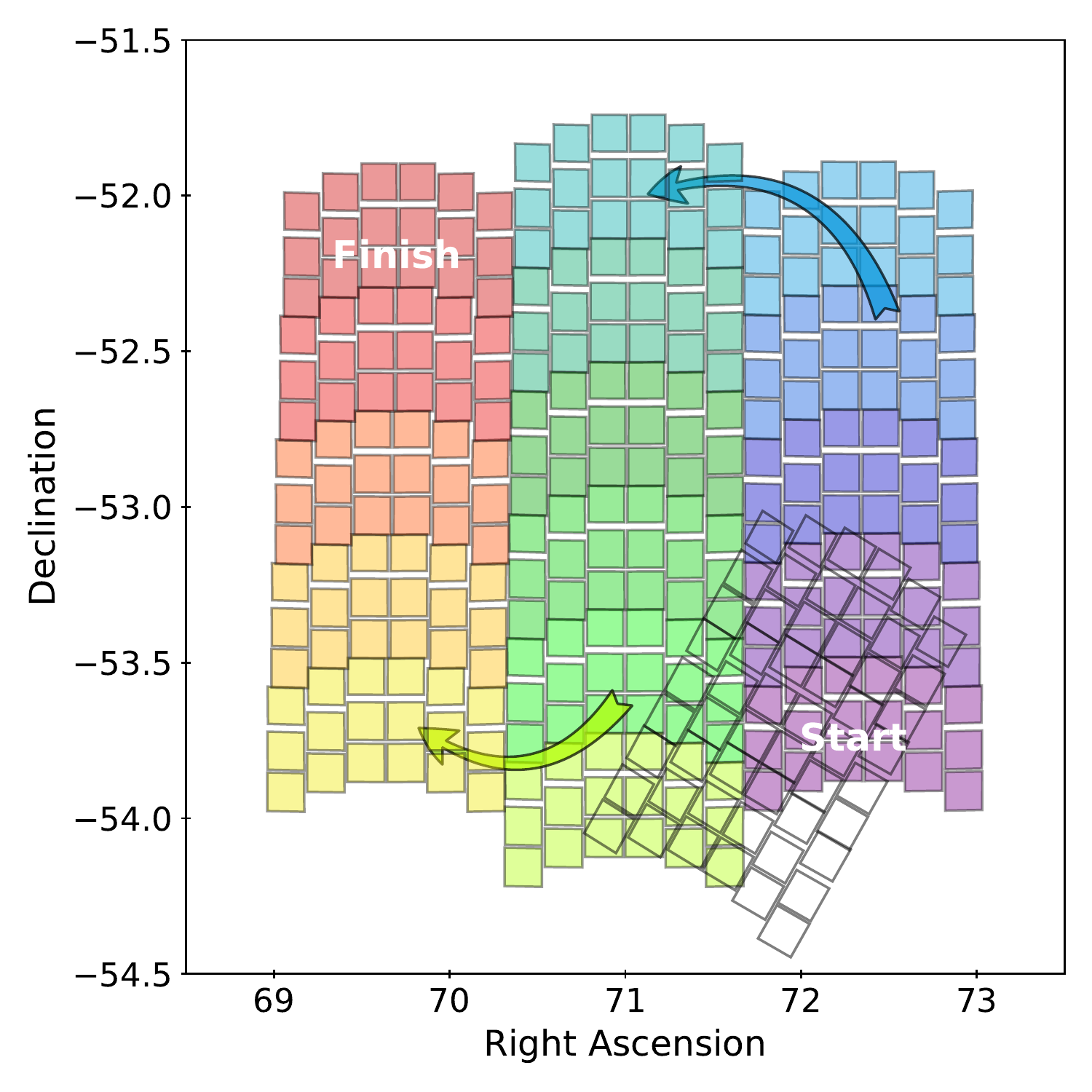}
    \caption{
    Possible slewing strategies. The left panel presents one full visit that embeds the deep tier (black pointing) inside the wide tier (gray). In total, this is a 12~deg$^2$ field, with expected slew and settle times indicated in the legend.  The right panel presents the basic ``snake plan'' slewing strategy for a 5~deg$^2$ deep tier. This sequence begins with the purple WFI footprint in the lower right, snaking to finish at the red position in the upper left, as indicated by the arrows. The uncolored footprints in the lower right show the first four pointings of a subsequent visit, rotated by 30~degrees.
    \label{fig:slew}
    }
\end{figure}

Both the \sn fields should be roughly circular so that the fields can be tiled in the same pattern as the observatory rotation angle changes. Finally, our recommendation is that the wide and deep portion of each field be concentric, such that if a \sn falls out of the deep survey due to the edge effects of tiling, then it lands in the wide survey instead of being missed.
We show that achieving this is possible with a slewing strategy dubbed the ``snake plan.'' 
A visualization of both the basic ``snake plan'' and one where the deep tier is embedded in the wide tier are presented in \cref{fig:slew}.

The basic ``snake plan'' was constructed by one central row of six pointings and two parallel rows of five pointings. Pointings are chosen to minimize gaps, with the inner circle as a $5~\text{deg}^2$ area and the outer circle, which is the largest extent of the chips, as $8~\text{deg}^2$.
After the whole pattern is observed, filters are switched, a chip gap dither is applied, and then the pattern is reversed.
No dithering is done while going through a single ``snake plan''.

The roll angle could either be the natural roll of the observatory ($\sim$1~$\text{deg}/\text{day}$) or $30~\text{deg}$ jumps to maintain a specific angle for as long as possible. This means either every \sn has 1/8th of its observations fall in a chip gap or a fraction of \sne lose a month of observations.
Studies are ongoing into which method has less impact on cosmological parameter inference.

The pattern of pointings described above can be scaled up or down to fit different field sizes. For the wide field, our plan (\cref{tab:25}) for the pointing structure consists of 57 pointings, the pattern is $11\times5$, with the center two columns being 6 pointings instead of 5. For the  spectroscopic wide and deep patterns, our plan is $4\times3$ and $2\times2$ respectively.

Prism observations will also be a rolling survey, participating in the above plan just like another broadband filter. However, for survey strategies with especially small prism areas (e.g., 2 or 4 pointings) active targeting with the prism may be more effective.

There are $\sim$100 pointings per visit. Specific values for pointings and area per tier are listed in columns 6 and 7 of \cref{tab:25}. These values are constrained by the exposure time (set by the target redshift) and time per visit.

\subsection{Exposure Times}

\begin{figure}
    \centering
    \includegraphics[width=\textwidth]{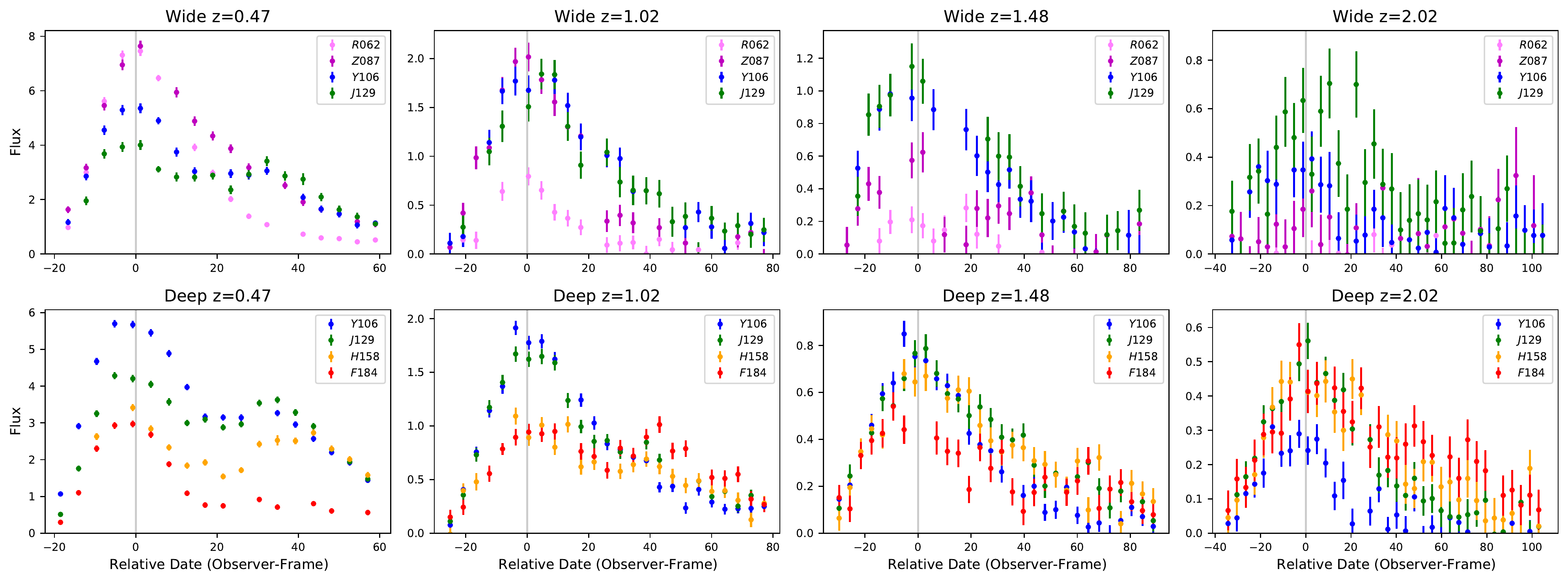}
    \caption{Simulated light curves as a function of redshift for the wide tier (top row) and the deep tier (bottom row). The \sn plotted for each redshift has the median total S/N (across all epochs and filters) of all \sne at that redshift.}
    \label{fig:SN_LCs}
\end{figure}

\begin{table}
\centering
\footnotesize
 \begin{tabular}{|l l c|c c r r r|r|} 
 \hline
 Mode & Tier & $z_{\mathrm{targ}}$\tablenotemark{*} & Filters & Exp.Time+Overhead & No. of & Area & Time/Visit & Total \\
  &  & &  & (s) & Pointings & (deg$^2$) & (hours) & \sn\\ [0.5ex] 
 \hline\hline
  \multicolumn{4}{l}{\textbf{25\% Spectroscopy Survey}}&\\\hline
 Imaging & Wide & 1.0 & RZYJ & 160;100;100;100 + 70x4 & 68 & 19.04 & 14.0 &8804 \\ 
 Imaging & Deep & 1.7 & YJHF & 300;300;300;900 + 70x4 & 15 &  4.20 & 8.5 &3520  \\ 
 \hline
 \textbf{Subtotal} & & &  &  &  &  & \textbf{22.5} & \textbf{12324} \\ 
 \hline
 Spec & Wide & 1.0 & prism & \hfill 900 + 70~~~~ & 12 & 3.36 & 3.2 & 831  \\ 
 Spec & Deep &  1.5 & prism & \hfill 3600 + 70~~~~ &  4 & 1.12 & 4.1 &652 \\ 
 \hline
 \textbf{Subtotal} & &  &  &  &  &  & \textbf{7.3} &  \textbf{1483}  \\ 
 \hline 
 \end{tabular}
\tablenotetext{*}{$z_{\mathrm{targ}}$ denotes the redshift where the average \sn at peak is observed with S/N=10 per exposure for imaging, and S/N=25 for spectroscopy.}
\caption{
A High-latitude Time Domain reference survey.
}
  \label{tab:25}
\end{table}

The exposure times for the imaging are set such that at a target redshift ($z_{\mathrm{targ}}$) a mean \sn at peak has ${\rm S/N} = 10$ per exposure. 
This S/N allows for precision photometry and light-curve modeling.
The prism exposure times were designed such that for a given target a S/N of 25 would be reached in the rest-frame V-band (assuming a synthetic top-hat) when co-adding spectra $\pm 5$ days (rest-frame) around peak. For redshifts $1 \leq z \leq 2$, these co-adds would have $\sim$5--7~epochs. The prism S/N is set such that the total light-curve S/N from the imaging and spectral time series are comparable. The overhead of a given exposure is roughly 70 seconds (slew plus settle); as such, our minimum exposure duration is 100 seconds.
The exposure times for the four filters, for each tier, are specified in \cref{tab:25}.
Example light curves for this strategy can be seen in \cref{fig:SN_LCs}.
A few alternatives to this survey---including a change in fraction of spectroscopic time and thus survey area---are presented in \cref{sec:alt}.

These imaging exposure times lead to a limiting magnitude of an isolated static point sources of $\sim$25.5~mag and $\sim$26.5~mag for the wide and deep tiers respectively.
We expect 144 epochs over the two-year survey. With a focal plane fill fraction of $\sim$87\% we expect $\sim$125 observations per object, resulting in deep co-adds having a limiting magnitude of 2.6~mag deeper than the individual exposures ($\sim$28~mag and $\sim$29~mag).
Specific limiting magnitudes for each filter can be found in \cref{tab:mag}.

For the prism exposures, the limiting redshift 
depends on analysis technique and what information the spectra are intended to supply (i.e., redshift, classification, standardization). Work is ongoing to quantify these limits.

\begin{table}
\centering
\footnotesize

\hspace{-0.5in} \begin{tabular}{| l |c c c c c c|} 
 \hline
 & F062/R & F087/Z & F106/Y & F129/J & F158/H & F184/F \\
\hline \hline
\multicolumn{2}{c}{\textbf{Wide Tier}} & \multicolumn{5}{c}{\phantom}\\
\hline
Exposure time (sec) & 160 & 100 & 100 & 100 & --- & --- \\
Single-exposure limiting magnitude & 26.4  & 25.6  & 25.5  & 25.4  & &\\
125-exposure co-add limiting magnitude & 29.0  & 28.2  & 28.1  & 28.0  & & \\
\hline
\multicolumn{2}{c}{\textbf{Deep Tier}}  & \multicolumn{5}{c}{\phantom}\\
\hline
Exposure time (sec) & --- & --- & 300 & 300 & 300 & 900 \\
Single-exposure limiting magnitude & & & 26.7  & 26.6  & 26.5  & 26.7\\
125-exposure co-add limiting magnitude & & & 29.3 & 29.2 & 29.1 & 29.3\\
\hline
\end{tabular}

  \caption{Limiting AB magnitudes for isolated point sources. Using a fill fraction of 87\% and 144 epochs over the two-year survey, the co-added depths are expected to be $\sim$2.6~mag deeper.}
  \label{tab:mag}
\end{table}

\subsection{Forecasts of the Number of \sne and Measuring $w_0$ \& $w_a$}\label{sec:numbers}

Catalog-level simulations and analysis, using the SNANA set of programs \citep{Kessler2009a,Kessler2010}, were performed to estimate the number of \sne that will be observed after basic detection requirements. This software requires characterizing various aspects of the survey, including filter transmission functions, catalogs of host galaxies, and a survey strategy. We detail non-\rst specific choices here.

For our catalog of potential \sn host galaxies, we follow the work done in Wang et~al.\ (in prep.) which utilizes the framework of \citet{Troxel2020} to map 38,000 real galaxies observed by the \textit{Hubble Space Telescope} (\textit{HST}) in the CANDELS program \citep{hemmati2019} to the much larger Buzzard simulation \citep{DeRose2019}. This simulation generates appropriate distributions for redshift, mass, clustering, and sky distribution, resulting in a host-galaxy library of $\sim$1.5~million~galaxies. 

Light curves are generated with the SALT2 model \citep[using a near-IR extension;][]{Guy2007, Pierel18}. Detection in a single filter requires ${\rm S/N} > 3$, and discovery/triggering requires detection in two bands (YJ for the wide tier and JH for the deep tier). In this analysis, we require \sne to be detected at least once within 5 days after peak, and again after 15 days (both in the rest frame). Though with a rolling survey with no seasons, these phase cuts alone remove very few \sne. For the imaging component, we also require that the host-galaxy redshift will be measured. Details of our estimated redshift efficiency are in preparation.  We follow the SN rate function described in \cite{Hounsell2018}.  In total, $\sim$200,000 \sne are simulated over the course of the survey. After all selection cuts, approximately 12,500 \sne pass through to the light-curve fitting stage. We note that alternative estimates of these numbers, outside the SNANA framework, agree to the level of $\sim$10\%.

\begin{table}
\centering
\footnotesize

 \begin{tabular}{|l |c c |r|} 
 \hline
Observational Cuts & \sn & SN Core Collapse & Total \\
\hline
Detections &  20,209 & 31,633 & 51,842 \\
$\geq 1$ observation before phase $+5$-days  & 19,155 & 29,998 & 49,153 \\
$\geq 1$ observation after phase $+15$-days &  17,616& 28,262  & 45,878 \\
$\geq 1$ observation with S/N $> 10$ &  12,427 & 14,419 & 26,846 \\
$\geq 2$ filters with an observation with S/N $> 5$ &  12,427 & 14,263 & 26,690 \\
 \hline
Total after light-curve fitting \& observational cuts &  12,324 & 147 & 12,471
 \\ 
\hline
\end{tabular}

  \caption{
Cuts applied and number of ``passing" SNe at the simulation and light-curve fitting stage for the 25\% spectroscopy survey (defined in \cref{tab:25}).}
  \label{tab:cuts}
\end{table}

\begin{figure}
    \centering
    \includegraphics[width=\linewidth]{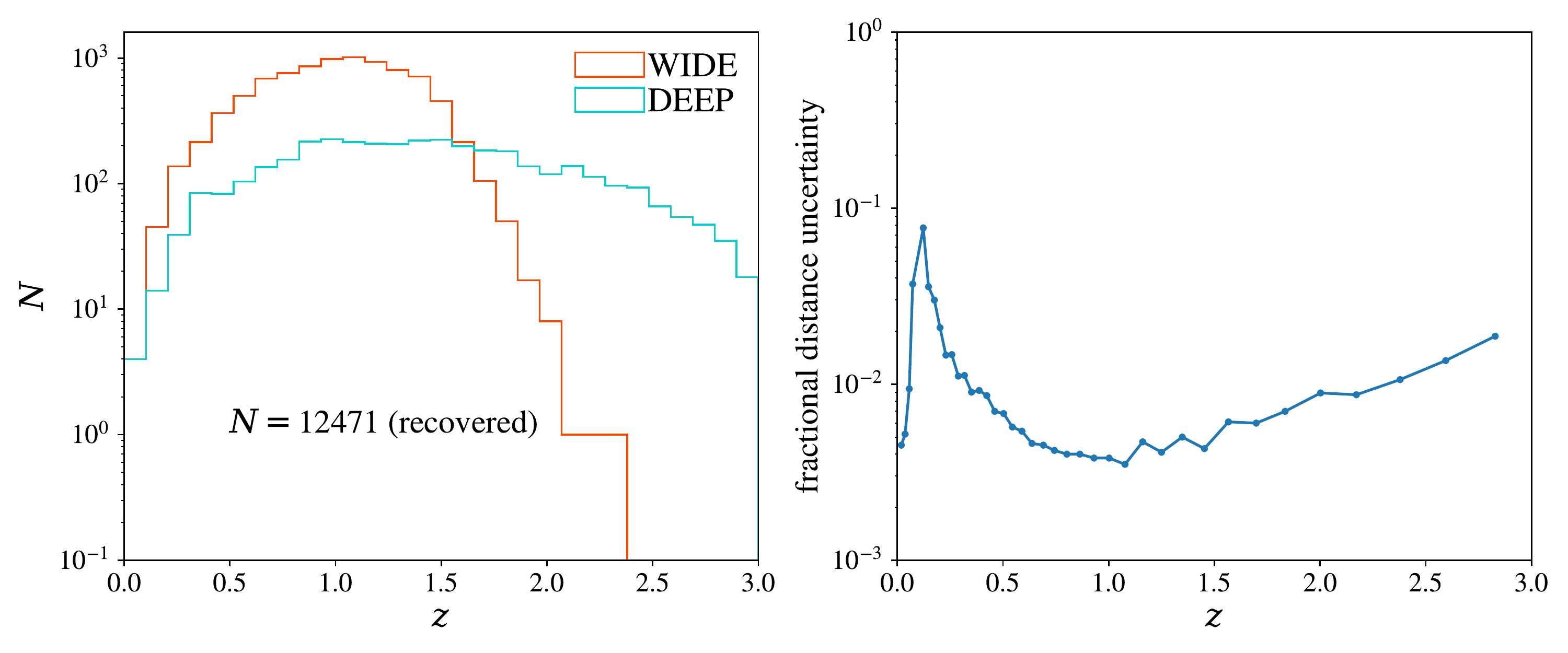} 
    \caption{\emph{Left}: Redshift histogram of SNe (Ia and core-collapse) which pass analysis cuts. We recover 8,861 in wide tier and 3,610 in deep tier, with 5,197 of the 12,471 total \sne coming from $z>1$. \emph{Right}: Fractional distance error using standardization with SALT2 light curves, counting statistical errors only, as a function of redshift after all quality-cuts are applied for our combined \rst + low-$z$ imaging sample. The increase in uncertainty around $z=0.1$ is due to the drop in number of \sne per redshift bin as the data set transitions between these two surveys.
    } \label{fig:recoveredhist}
\end{figure}

For light-curve fitting, we again utilize the SALT2 model to calculate best-fit color and stretch parameters. 
We also apply several cuts here, including a ${\rm S/N} > 10$ obtained in at least one filter and ${\rm S/N} > 5$ in at least two filters. Cuts and the number of \sne passing each of these cuts are detailed in Table \ref{tab:cuts}. 
With these cuts applied, the number of \sne expected are listed in \cref{tab:25}. For this survey, 12,324 \sne (plus 147 core-collapse interlopers) are kept for cosmological analysis. 
In total, 8,861 and 3,610 SN in the wide and deep tiers, respectively, pass observational cuts, with 5,197 of the 12,471 total \sne at $z>1$.
The redshift distribution of passing \sne is shown in the left panel of \cref{fig:recoveredhist}. Also in \cref{fig:recoveredhist}, we present the statistical distance uncertainty as a function of redshift based on standardization using the SALT2 light curve parameters, for comparison with Figure 3-5 of the \rst Science Requirements Document.

We can check whether the numbers shown here would satisfy the requirements of a $\sim$1\% measurement in $w_0$ and a $\sim$10\% measurement in $w_a$.  To do so, one must measure a deviation from $\Lambda$CDM on the order of 6~mmag from $z=0$ to $z=2$ (the redshift range including a low-$z$ sample) or $\sim$3~mmag from $z=0.5$ to $z=2$ (the effective redshift range of \rst only) To do so requires roughly 4~bins across the \romanST range with a precision of 3~mmag.  Given a distance modulus precision from imaging of roughly 0.1~mag \citep{Hounsell2018}, one needs 2500~\sne in a redshift bin to reach this level. Similarly, given a distance modulus precision from spectroscopy of roughly 0.06~mag \citep{Fakhouri2015,Boone2021a,Boone2021b}, one needs 900~\sne in a redshift bin to reach this level. Multiplying each of these numbers by 4~bins, the total needed for imaging and spectroscopy would be $\sim$10,000 and $\sim$3,600 respectively. These numbers can be satisfied with either a $75\%$ imaging or $75\%$ spectroscopy time survey. Therefore, we find the cosmological goals can be achieved with the survey presented here along with a set of other strategies (see \cref{sec:alt}), though with little margin for survey decrease.

\subsection{Example Alternative Strategies}\label{sec:alt}

There are many ways to adjust a survey, but most will have a minor impact on final data products and analysis methods.
The biggest effect would be a change in the fraction of time dedicated to spectroscopy, resulting in a change in the area surveyed, how the analysis will be performed, and what tools need to be built.
In addition, a shift of emphasis from the rest-frame optical to the NIR would change exposure times, filters, and areas, as well as requiring new analysis techniques.
Changing the number of tiers, or the cadence, would also affect the total observing area. 
Each of these changes will also impact the systematic uncertainties in complicated ways that require detailed study.
As an example case, here we examine alternative fractions of spectroscopic observing times.

There will be a number of drivers influencing the eventual fraction of \sn for which we will obtain prism spectroscopy, including: the relative need for spectroscopic host-galaxy redshifts, SN classification, and \sn evolution indicators. We here describe options for the fraction of survey time dedicated to prism spectroscopy, ranging from 10\% to 75\%, as summarized in \cref{tab:alt}. In this note, we began the discussion by describing the case in which  25\% of the time is dedicated to prism spectroscopy, we consider this our nominal plan to be used for initial future simulations. These surveys are based on our current understanding of the aforementioned issues regarding spectroscopy in addition to our understanding of the systematic errors i.e.,  this is a work in progress.

\begin{table}[]
\centering
\footnotesize
 \begin{tabular}{|l l c|c c r r r|r|} 
 \hline
Mode & Tier & $z_{\mathrm{targ}}$\tablenotemark{*} & Filters & Exp.Time+Overhead & No. of & Area & Time/Visit & Total\\

  &  & &  & (s) & Pointings  & (deg$^2$) & (hours) &  \sn\\ [0.5ex] 
 \hline\hline
 \multicolumn{9}{l}{\phantom}\\
 \multicolumn{3}{l}{\textbf{10\% Spectroscopy Survey}} & \multicolumn{5}{c}{\phantom}\\
 \hline

 Imaging & Wide & 1.0 & RZYJ & 160;100;100;100 + 70x4 & 82 & 22.96 & 16.8 & 10617 \\ 
 Imaging & Deep & 1.7 & YJHF & 300;300;300;900 + 70x4 & 18 &  5.04 & 10.2 & 4224 \\ 
 \hline
 \textbf{Subtotal} &  &  &  &  &  &  & \textbf{27.0} & \textbf{14841} \\ 

 \hline
 Spec & Wide & 1.0 & prism & \hfill 900 + 70~~~~ & 4 & 1.12 &  1.0 & 277 \\ 
 Spec & Deep & 1.5 & prism & \hfill 3600 + 70~~~~ &  2 & 0.56 &  2.0 & 326 \\ 
 \hline
 \textbf{Subtotal} & &  &  &  &  &  & \textbf{3.0} & \textbf{603} \\ 
 \hline 
 \multicolumn{9}{l}{\phantom}\\
 \multicolumn{3}{l}{\textbf{ 50\% Spectroscopy Survey}}&\multicolumn{5}{c}{\phantom}\\

 \hline
 Imaging & Wide & 1.0 & RZYJ & 160;100;100;100 + 70x4 & 45 & 12.60 & 9.3 & 5826 \\ 
 Imaging & Deep & 1.7 &   YJHF & 300;300;300;900 + 70x4 & 10 &  2.80 & 5.8 & 2347 \\ 
 \hline
 \textbf{Subtotal} &  &  &  &  &  &  & \textbf{15.1} & \textbf{8173} \\ 
 \hline
 Spec & Wide & 1.0 & prism & \hfill 900 + 70~~~~ & 25 & 7.00 & 6.7 & 1731 \\ 
 Spec & Deep & 1.5 & prism & \hfill 3600 + 70~~~~ &  8 & 2.24 & 8.2 & 1302 \\ 
 \hline
 \textbf{Subtotal} & &  &  &  &  &  & \textbf{14.9} & \textbf{3032}
\\ 
\hline
 \multicolumn{9}{l}{\phantom}\\
  \multicolumn{3}{l}{\textbf{ 75\% Spectroscopy Survey}}&\multicolumn{5}{c}{\phantom}\\
  \hline
  Imaging & Wide & 1.0 & RZYJ & 160;100;100;100 + 70x4 & 19 & 5.32 & 3.9 & 2460 \\ 
  Imaging & Deep & 1.7 & YJHF & 300;300;300;900 + 70x4 & 6 & 1.68 & 3.5 & 1408 \\ 
  \hline
  \textbf{Subtotal} &  &  &  &  &  &  & \textbf{7.4} & \textbf{3868} \\ 
  \hline
  Spec & Wide & 1.0 & prism & \hfill 900 + 70~~~~ & 19 & 5.32 &  5.1 & 2460 \\ 
  Spec & Deep & 1.7 & prism & \hfill 10400 + 70~~~~ &  6 & 1.68 & 17.5 & 1408 \\ 
  \hline
  \textbf{Subtotal} & &  &  &  &  &  & \textbf{22.6} & \textbf{3868} \\ 
\hline
  \end{tabular}
 \tablenotetext{*}{$z_{\mathrm{targ}}$ denotes the redshift where the average \sn is  observed with S/N=10 per exposure.}
 \caption{Alternative reference surveys. The key difference between these surveys is the number of pointing and how the time is split between imaging and spectroscopy. The 75\% strategy also increases the target spectroscopic depth.} \label{tab:alt}
\end{table}

When 75\% of the time is devoted to prism spectroscopy, we impose the condition that the area for spectroscopy can be no larger than the imaging area. As such, we get spectra only for those \sne that have light curves. Therefore, in this particular formulation of the survey, the advantage of a 75\% prism-spectroscopy-focused survey are deeper spectra (reaching higher redshift) rather than more \sne with spectra. In this case, the imaging and spectroscopic areas would be equal, so all \sne in the Hubble-Lema\^itre Diagram would have multi-epoch spectroscopy. However, there are other alternatives possible for a 75\% survey, which will be investigated in the future.

We also note that the survey does not need to have one set strategy for its entire duration. One helpful variation would be that the survey may start with a higher fraction of prism time, say 50\%, for the first few months. This ``top heavy'' approach would produce a sufficient spectroscopic sample for an early look at the potential evolution of \sne, and determination of accurate IR-SN spectral-temporal templates. Additionally, setting aside some fraction of prism time for host-galaxy redshifts at the end of the survey could allow a more targeted strategy to boost the final sample size. Depending on the results obtained in a pre-survey period, the fraction of time devoted to spectroscopy could be altered as appropriate.

\section{Conclusions}

The \romanFull 
will be able to conduct a generation-defining \sn cosmology analysis and fulfill all of its mission requirements for such an experiment. We expect the \romanST to be able to measure a $\sim$6~mmag deviation from $\Lambda$CDM over a redshift range from $z=0$ to $z=2$. 

This note describes a survey strategy that use six filters and the prism on the \rst Wide Field Instrument. The survey has two tiers, one ``wide'' which targets \sne at redshifts of $z=1$ and one ``deep'' targeting $z=1.7$; for each, 4 filters are used to cover the rest-frame optical wavelength range of these redshifts.  The tiers will be observed  in at least two separate fields to both reduce cosmic variance biases and provide overlap with external programs.  We propose one field each in the north and south continuous viewing zones, and expect to obtain observations of $\sim$12,000 \sne depending on the total area and depth of the survey. Exposure times range from $100~\text{s}$ to $900~\text{s}$ for imaging and $900~\text{s}$ to 3,$600~\text{s}$ for the prism. Each visit would have, $\sim$100 pointings, and a  cadence of $\sim$5 days between sets of pointings. The total survey spans two years with a total survey time of six months.
Though consensus has not been achieved regarding a single observing strategy, we consider this survey a reasonable starting point to be used in future simulations.

\acknowledgments
The Supernova Science Investigation Teams are supported by NASA through grants NNG16PJ311I, NNG17PX03C and NNG16PJ34C.

This work was completed in part with resources provided by the University of Chicago’s Research Computing Center.

This material is based upon work supported by NASA under award number 80GSFC21M0002.
The UCSC team is additionally supported in part by NASA grant 14-WPS14-0048, NSF grants AST-1518052 and AST-1815935, the Gordon and Betty Moore Foundation, the Heising-Simons Foundation, and by fellowships from the Alfred P.\ Sloan Foundation and the David and Lucile Packard Foundation to R.J.F.

\software{
Astropy \citep{Astropy},
\texttt{COSMOMC} \citet{Lewis2013},
emcee \citep{Foreman-Mackey2013,Foreman-Mackey2019},
Matplotlib \citep{matplotlib},
NumPy \citep{numpy},
Python,
pyLINEAR \citep{Ryan2018},
pSNid \citep{Sako2011},
Seaborn (\doi{10.5281/zenodo.592845}),
SNCosmo \citep{sncosmo},
SNANA \citep{Kessler2009a,Kessler17},
}

\bibliographystyle{aasjournal}
\bibliography{library}

\allauthors

\end{document}